\documentclass[12pt,a4paper,dvips]{article}
\usepackage{graphicx}
\usepackage{hhline}
\usepackage{amsmath}
\usepackage{amssymb}
\usepackage{times}
\usepackage[varg]{txfonts}
\usepackage{array}
\usepackage{multirow}
\usepackage{dcolumn}

\usepackage{caption2}

\newcommand{\tablecaption}{%
\setlength{\abovecaptionskip}{0pt}
\setlength{\belowcaptionskip}{10pt}
\caption}

\newcolumntype{.}{D{.}{.}{-1}}
\newcolumntype{-}{D{-}{-}{-1}}

\usepackage{color}
\definecolor{rltred}{rgb}{0.75,0,0}
\definecolor{rltgreen}{rgb}{0,0.5,0}
\definecolor{rltblue}{rgb}{0,0,0.5}

\usepackage[numbers,square,comma,sort&compress]{natbib}
\usepackage{hypernat}
\newcounter{pdfadd}    
\usepackage[hyperindex,bookmarks,bookmarksnumbered,breaklinks,a4paper,unicode]{hyperref}
\hypersetup{%
  pdftitle        = {Measurement of the Proton Structure Function F2 at low Q2 in QED Compton Scattering at HERA},
  urlcolor        = rltblue,       
  urlbordercolor  = 0 0 0.5,
  filecolor       = rltblue,       
  filebordercolor = 0 0 0.5,
  linkcolor       = rltred,        
  linkbordercolor = 0.75 0 0,
  citecolor       = rltgreen,      
  citebordercolor = 0 0.5 0,
  pagecolor       = rltgreen,      
  pagebordercolor = 0 0.5 0,
  menucolor       = rltgreen,      
  menubordercolor = 0 0.5 0,
  pdfauthor     = {H1 Collaboration},
  pdfsubject    = { },
  pdfkeywords   = {High-Energy Physics, Particle Physics, Proton Structure}
}

\newlength{\dinwidth}
\newlength{\dinmargin}
\def\gsim{\,\lower.25ex\hbox{$\scriptstyle\sim$}\kern-1.30ex%
\raise 0.55ex\hbox{$\scriptstyle >$}\,}
\def\lsim{\,\lower.25ex\hbox{$\scriptstyle\sim$}\kern-1.30ex%
\raise 0.55ex\hbox{$\scriptstyle <$}\,}

\begin{document}  

\makeatletter \def\NAT@space{} \makeatother

\begin{titlepage}

\noindent

DESY 04--083 \hfill ISSN 0918--9833\\

\vspace*{3.5cm}

\begin{center}
\begin{Large}

{\bfseries 
    Measurement of the Proton Structure Function {\boldmath $F_2$} \\ 
    at low {\boldmath $Q^2$} in QED Compton Scattering at HERA}

\vspace{2cm}

H1 Collaboration

\end{Large}
\end{center}

\vspace{2cm}

\begin{abstract}
\noindent
The proton structure function $F_2(x,Q^2)$ is measured in inelastic QED Compton
scattering using data collected with the H1 detector at HERA.
QED Compton events are used to access the kinematic range of very low
virtualities of the exchanged photon, $Q^2$, down to $0.5\,{\rm GeV}^2$,
and Bjorken $x$ up to $\sim$$0.06$, a region which has not been
covered previously by inclusive measurements at HERA. The results are in
agreement with the measurements from fixed target lepton-nucleon scattering
experiments.
\vspace{1.5cm}
\end{abstract}

\begin{center}
{\slshape Submitted to Physics Letters B}
\end{center}

\end{titlepage}

\begin{flushleft}

A.~Aktas$^{10}$,               
V.~Andreev$^{26}$,             
T.~Anthonis$^{4}$,             
A.~Asmone$^{33}$,              
A.~Babaev$^{25}$,              
S.~Backovic$^{37}$,            
J.~B\"ahr$^{37}$,              
P.~Baranov$^{26}$,             
E.~Barrelet$^{30}$,            
W.~Bartel$^{10}$
% ...

\bigskip{\slshape 
% ...
$ ^{4}$ Inter-University Institute for High Energies ULB-VUB, Brussels;
          Universiteit Antwerpen, Antwerpen; Belgium$^{ c}$ \\
% ...
$ ^{10}$ DESY, Hamburg, Germany \\
% ...
$ ^{25}$ Institute for Theoretical and Experimental Physics,
          Moscow, Russia$^{ l}$ \\
 $ ^{26}$ Lebedev Physical Institute, Moscow, Russia$^{ e}$ \\
% ...
$ ^{30}$ LPNHE, Universit\'{e}s Paris VI and VII, IN2P3-CNRS,
          Paris, France \\
% ...
$ ^{33}$ Dipartimento di Fisica Universit\`a di Roma Tre
          and INFN Roma~3, Roma, Italy \\
% ...
$ ^{37}$ DESY, Zeuthen, Germany \\
% ...
$ ^c$ Supported by FNRS-FWO-Vlaanderen, IISN-IIKW and IWT
      and  by Interuniversity Attraction Poles Programme,
      Belgian Science Policy \\
% ...
$ ^e$ Supported by the Deutsche Forschungsgemeinschaft \\
% ...
$ ^l$ Partially Supported by Russian Foundation
      for Basic Research, grant    no. 00-15-96584 \\
}

\end{flushleft}

\newpage

\section{Introduction} \label{s:intro}
Measurements of deep inelastic lepton-nucleon scattering (DIS) provide
information which is crucial to our understanding of proton structure and 
which has played a decisive role in the development 
of the theory of strong interactions, Quantum Chromodynamics (QCD).
Since the discovery of Bjorken scaling~\cite{Bloom:1969kc}
and its violation~\cite{Fox:1974ry} at fixed target experiments,
much progress has been made in extending the kinematic range of measurements
in terms of the Bjorken variable $x$ and the modulus of four-momentum transfer
squared $Q^{2}$. The H1 and ZEUS experiments at the HERA {\itshape ep} collider
have shown that the
$Q^{2}$ evolution of the proton structure function $F_2(x,Q^2)$ is well 
described by perturbative QCD (pQCD) throughout a wide range in $x$ and
$Q^{2}$~\cite{Adloff:2000qk,Adloff:2003uh,Chekanov:2001qu,Chekanov:2003yv}.
However, at small $Q^2$, where the transition takes place into a region
in which non-perturbative effects dominate, pQCD is no longer applicable.
The data in this region~\cite{Adloff:1997mf,Breitweg:2000yn} are described
by phenomenological models such as those derived from the Regge approach.

In order to study this non-perturbative regime, the structure function
$F_2$ has been measured in {\itshape ep} scattering at HERA for very low
values of $Q^{2}$ using a detector mounted close to the outgoing 
electron\footnote{The generic name ``electron'' is used to denote both
electrons and positrons.} beam direction~\cite{Breitweg:2000yn}.
The transition region around \mbox{$Q^{2}$ $\sim 1$\,GeV$^{2}$}
has been investigated using data taken in dedicated runs with the
interaction vertex shifted~\cite{Derrick:1996ef,Adloff:1997mf}.
In this paper a new complementary measurement of $F_2$ in this kinematic domain
is presented, which, following the discussion in~\cite{Lendermann:2003rq},
utilises {\itshape ep} data with wide angle hard photon radiation,
so called QED Compton events.

The data used in this analysis correspond to a luminosity of 
$9.25\,{\rm pb}^{-1}$ and cover the kinematic range of $Q^2$
between $0.5$ and $7$\,GeV$^2$. As compared with previous HERA
analyses, the range of Bjorken $x$ is significantly extended towards rather
large values, between $0.001$ and $0.06$. This is achieved
by introducing a detailed simulation of the hadronic final
state at low masses $W$ which includes the resonance region,
and through an improved understanding of calorimeter noise.

\section{QED Compton Scattering Cross Section} \label{s:theory}
\begin{figure}[tb]
\centerline{\includegraphics[width=.7\textwidth]{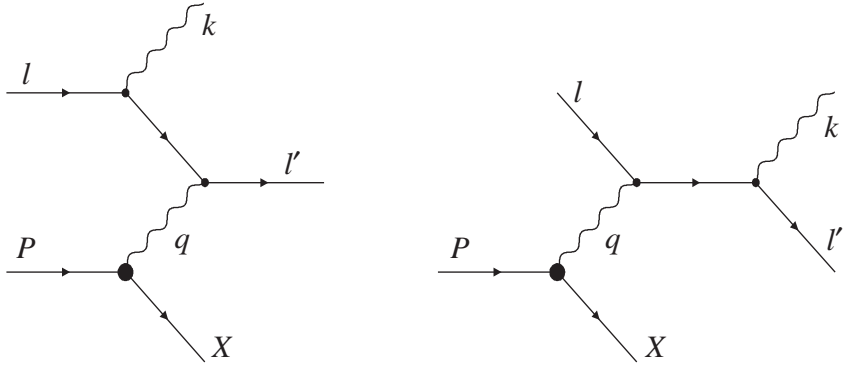}}
\caption{Lowest order Feynman diagrams for the radiative process
$ep \rightarrow e \gamma X$ with photon emission from the electron line.
Here $l$ and $P$ represent the four-momenta of the incoming electron and the 
incoming proton, while $l^{\prime}$, $k$ and $X$ are the four-momenta of 
the scattered electron, the radiated photon and the hadronic final 
state, respectively.}
\label{f:feynman_rad}
\end{figure}

Radiative processes in {\itshape ep} scattering, as depicted in
Fig.\,\ref{f:feynman_rad}, are of special interest,
since the photon emission from the lepton line gives rise to event kinematics
which open new ways of investigating proton structure~%
\cite{Krasny:1991hd,Blumlein:1993ef,Ahmed:1995cf,Derrick:1996ef,Aid:1996au,DeRujula:1999yq,Lendermann:2003rq}.
In the present analysis the QED Compton (QEDC) process is considered, which is
characterised by low virtuality of the exchanged photon and high virtuality
of the exchanged electron. The experimental signature is an approximately 
back-to-back azimuthal configuration of the outgoing electron and photon.
In this configuration it is possible to reconstruct the event kinematics
for cases where the exchanged photon virtuality $Q^2$ is very small.

To describe the process $ep\rightarrow e\gamma X$ the standard
Lorentz-invariant kinematic variables $x$ and $Q^2$
have to be defined in a manner which accounts for the additional photon 
in the final state:
\begin{equation}
Q^2 = -q^2 = -(l - l^{\prime} - k)^2 ~, ~~~~~~~
x   = \frac{Q^2}{2 P \cdot (l - l^{\prime} - k)} ~.
\label{eq:kin_e}
\end{equation}
Here $l$ and $P$ are the four-momenta of the incoming electron and the
incoming proton, while $l^{\prime}$ and $k$ represent the four-momenta of the
scattered electron and the radiated photon, respectively
(Fig.\,\ref{f:feynman_rad}).
These variables can also be calculated in the usual way using the
four-momentum of the hadronic final state.
Neglecting the particle masses, the inelasticity $y$ used for the data
treatment can be obtained from the relation
$Q^2 = x y s$, where $s = (l + P)^2$ is the {\itshape ep}
centre-of-mass energy squared.

A complete calculation of the QEDC scattering cross section can be found
in~\cite{Courau:1992ht}.
Depending on the value of the invariant mass of the hadronic final state,
$W = [Q^2 (1 - x) / x + m_p^2]^{\frac{1}{2}}$, $m_p$ being the proton mass,
three separate contributions to the cross section are considered:
\begin{enumerate}
 \item {\bfseries Elastic scattering}, in which the proton stays intact
 ($W = m_{p}$). The cross section is calculated from the electric and
 magnetic form factors of the proton;
 \item {\bfseries Resonance production}, where the total mass of the
 hadronic final state $X$ lies in the range
 $m_p < W \lesssim 2$\,GeV;
 \item {\bfseries Continuum inelastic scattering} at $W \gtrsim 2$\,GeV. 
 In this region the cross section is
 defined through the proton structure functions $F_2$ and $F_L$.
\end{enumerate}
The dependence on the longitudinal structure function $F_L$ can be neglected
in the kinematic range studied in the present analysis, such that the cross
section is proportional to $F_2$.
The continuum inelastic QEDC events are thus used
to determine the structure function $F_2(x,Q^2)$.

\section{Experimental Technique} \label{s:detector}
The analysed events were recorded with the H1 detector~\cite{Abt:1997hi},
which consists of a number of subdetectors
designed to perform complementary measurements of particles created in high
energy {\itshape ep} collisions. The outgoing electron and photon in QEDC
events are selected by requiring two energy depositions (clusters) in the
electromagnetic section of the backward\footnote{The
$z$ axis of the right-handed coordinate system used by H1 is defined by
the direction of the incident proton beam with the origin at
the nominal {\itshape ep} interaction vertex.
Consequently, small scattering angles of the final state electron and photon
correspond to large polar angles in the H1 coordinate system.} 
lead-fibre calorimeter SpaCal~\cite{Appuhn:1996vf}.
The SpaCal covers the polar angle range of $153^{\circ} < \theta < 177^{\circ}$
and has an electromagnetic energy resolution of
$\sigma_E / E = 7\% / \sqrt{E/{\rm GeV}} \oplus 1\%$.
The final calibration of the electron and photon cluster energies is performed
using the double angle reconstruction method~\cite{Bentvelsen:1992} with
elastic QEDC events~\cite{Ahmed:1995cf,Lendermann:2002}.

The interaction vertex, necessary for polar angle reconstruction,
is determined from the intersection of the beam axis with an electron track
segment reconstructed in the backward region of the detector.
In addition to the electron, the final state photon can also leave a track,
if it converts into an electron-positron pair while traversing the tracking
chambers.
In events in which the electron or the converted photon is scattered into the
inner part of the SpaCal, the Backward Silicon Tracker (BST)~\cite{Eick:1996gv}
is used to reconstruct the vertex position. The BST has an angular coverage 
of $171.5^{\circ} < \theta < 176.5^{\circ}$ and a $\theta$ resolution of 
$0.3$\,mrad.
For polar angles $\theta \lesssim 172^{\circ}$ outside the BST acceptance
the Central Inner Proportional Chamber (CIP)~\cite{Abt:1997hi} is employed
in conjunction with the Backward Drift Chamber (BDC)~\cite{Schwab:1996}
or the SpaCal to determine the vertex coordinates.
The BDC is situated in front of the SpaCal and covers a similar angular range.
The polar angle resolution for this method of track reconstruction
varies between $1.3$ and $2$\,mrad.

For the reconstruction of the kinematic variables $Q^2$, $x$ and $y$
it is not necessary to distinguish experimentally between the outgoing electron
and photon, since the four-momenta of the two particles appear symmetrically
in eq.~(\ref{eq:kin_e}).
However, at least one of the particles is required to leave a signal in the
tracking chambers BST or CIP, in order to determine the interaction vertex
coordinates.
If both particles produce tracks, the vertex reconstructed with the 
smaller uncertainty is chosen.

The track reconstruction efficiencies are determined using inclusive DIS data 
and are found to be $91\%$ on average for the BST and ${\gtrsim}\,99\%$ for
the CIP.
Due to photon conversions some of the events for which the electron track
is not reconstructed remain in the sample.
The photon conversion rates amount on average to $21\%$ for the BST and
$17\%$ for the CIP acceptance range.

Inelastic QEDC events are distinguished from elastic events by demanding
hadronic activity in the detector.
The hadronic energy flow is measured in the Liquid Argon Calorimeter
(LAr)~\cite{Abt:1997hi} covering the angular range 
$4^{\circ} < \theta < 153^{\circ}$. 
Its hadronic energy resolution, as determined in test beam measurements~%
\cite{Andrieu:1993tz}, is about 
$\sigma_E / E = 50\% / \sqrt{E/{\rm GeV}} \oplus 2\%$.

The luminosity is measured using low angle bremsstrahlung events by tagging
the photon in a photon detector located at $z = -103\,{\rm m}$.

The detector acceptance and background contributions are calculated using
Monte Carlo simulations, as described in the following sections.
The detector response for events generated by the MC programs is simulated
in detail by a program based on GEANT3~\cite{Geant:1994}.
The simulated events are subject to the same reconstruction procedure
as the data.

\section{QEDC Event Simulation} \label{s:simulation}
QEDC events are simulated by the COMPTON Monte Carlo event generator~%
\cite{Courau:1992ht,Carli:1991yn}, which incorporates a complete calculation
of the leading order QEDC cross section.
The program also includes higher order corrections for Initial State
Radiation in the peaking approximation~\cite{Etim:1967}. 
For the investigation of inelastic QEDC events an improved version of the
COMPTON generator was developed~\cite{Lendermann:2002} which includes detailed
parameterisations for the resonance
(Brasse {\itshape et al.}~\cite{Brasse:1976bf}) 
and the continuum (ALLM97~\cite{Abramowicz:1997ms}) regions.
For the simulation of the hadronic final state the SOPHIA
program~\cite{Mucke:2000yb} is used in the range of low $Q^2$
($Q^2 < 2\,{\rm GeV}^2$) or low masses $W$ ($W < 5\,{\rm GeV}$). At higher $W$
and higher $Q^2$ the Quark Parton Model with subsequent Lund string
fragmentation \cite{Sjostrand:2000wi} is employed.

The SOPHIA model
provides an accurate description of photon-hadron interactions reproducing
a large set of available data~\cite{Mucke:2000yb}.
The simulation includes the production of the major baryon resonances,
direct pion production, multiparticle production based on the
Dual Parton Model~\cite{Capella:1994yb} with subsequent Lund string 
fragmentation, as well as 
the diffractive production of the light vector mesons $\rho$ and $\omega$.

\section{Background Treatment} \label{s:background}
A prominent background to inelastic QEDC scattering arises from inclusive
DIS events in which one particle from the hadronic final state
(typically a $\pi^0$) fakes the outgoing photon.
At high $y$, where the hadronic final state lies mostly in the backward region,
this process dominates the QEDC signature, hampering a clean QEDC event
selection. For this reason the measurement is restricted to relatively low $y$
values (see Sect.~\ref{s:selection}).
Remaining background from inclusive DIS events is modelled using the DJANGO MC
generator~\cite{Schuler:1991}. This includes LEPTO~\cite{Ingelman:1991} and
ARIADNE~\cite{Lonnblad:1992tz} for the hard interaction and higher order QCD
effects, as well as HERACLES~\cite{Kwiatkowski:1992es} for the calculation of
leading order QED radiative corrections.
In order to avoid double counting, DJANGO events with hard photons emitted
from the lepton line are excluded from the analysis if they fall into the 
phase space covered by the COMPTON generator.

{\itshape A priori} it cannot be expected that the specific non-perturbative
process of isolated \mbox{(pseudo-)} scalar meson production is correctly
simulated by an inclusive event generator.
Therefore a dedicated study of the DIS background was performed in which the 
inelastic QEDC selection is extended to events with three electromagnetic
SpaCal clusters.
One cluster is associated to the scattered electron by matching it to a BST
or CIP track segment. Decays of the pseudoscalar mesons $\pi^0$ or $\eta$ 
into two photons\footnote{The $\eta$ contribution to the background is small.
It is used for this study due to the larger opening angles between the photons
in the $\eta \rightarrow \gamma\gamma$ decays, compared with the $\pi^0$
decays.}
are selected using the invariant mass of the two other clusters. In this study
DJANGO is found to provide a reasonable description of events containing
$\pi^0$ or $\eta$ mesons.
The DIS background is estimated to contribute up to $12\%$ to the total
cross section with a systematic uncertainty of $50\%$ of the contribution.

Another source of significant background is Deeply Virtual Compton 
Scattering (DVCS), in which the final state photon is diffractively produced
in the virtual photon proton collision. DVCS and QEDC are indistinguishable on
an event-by-event basis, but differ in the kinematic distributions of the
outgoing electron and photon. Both processes can be simulated separately,
as the interference between them does not influence the energy and polar angle
distributions of the final state particles in the leading twist approximation.

Elastic DVCS events are simulated by the TINTIN generator~\cite{Stamen:2001}.
The cross section is normalised to the H1 results~\cite{Adloff:2001cn}.
The elastic DVCS channel contributes to the measured inelastic QEDC cross
section only if noise in the LAr calorimeter is misidentified as hadronic
activity. The admixture of such events is negligible.

A sizeable background to the inelastic QEDC cross section arises from
proton-dissociative DVCS, which has not yet been measured. In order to
estimate the size of this background, it is assumed that the diffractive
vertices factorise, such that the ratio of the DVCS cross section with
proton dissociation to that with an elastically scattered proton can be
modelled using the same ratio for $\rho^0$
electroproduction~\cite{Adloff:1997jd}. This assumption has been checked
in~\cite{Stamen:2001}. The inelastic DVCS process is then
estimated to contribute $5.5\%$ to the measured cross section and a $100\%$
uncertainty is assigned to the resulting background subtraction.

Further background sources considered are:
\begin{itemize}
\item elastic QEDC events, contaminated by electronic noise in the LAr
calorimeter. These events contribute $0 - 2$\% to the measured signal;
\item elastic and inelastic dielectron production, modelled using the
GRAPE event generator~\cite{Abe:1999}. The contribution varies between
$0.5$ and $2\%$;
\item inclusive photoproduction, simulated by the PHOJET program
\cite{Engel:1996yd}, which in particular includes the production of the light
vector mesons $\rho$, $\omega$ and $\phi$.
The contribution is $\lesssim 0.5\%$;
\item diffractive electroproduction of light vector mesons as well as
diffractive $J/\psi$ photo- and electroproduction, all being simulated
by the DIFFVM MC generator~\cite{List:1999} 
for both the elastic and the proton-dissociative case. Sizeable backgrounds
arise from $J/\psi$ photoproduction ($\lesssim 3\%$) and from
$\rho$ electroproduction ($\lesssim 1.5\%$);
\item two photon resonance production. The contribution of this process
is estimated analytically employing results
from~\cite{Kilian:1998ew} and is found to be negligible;
\item beam-induced background, estimated using non-colliding 
particle bunches. This contribution is found to be below $0.5\%$.
\end{itemize}

\section{Event Selection} \label{s:selection}
The measurement is performed using $e^+ p$ data with a centre-of-mass energy of
$\sqrt{s} = 301$\,GeV. The data were recorded in 1997 and correspond to an
integrated luminosity of $9.25\,{\rm pb}^{-1}$.

The analysed events were collected by combining two trigger arrangements.
The first selects events with two clusters in the electromagnetic part
of the SpaCal. This trigger excludes the inner region of the SpaCal,
as event rates become large at low scattering angles.
The event selection based on this trigger is performed only in
a fiducial region in which it is fully efficient.
For the present measurement $59\%$ of the events are
collected by this trigger.

The remaining events were selected using another trigger arrangement,
which selects events with two clusters in the electromagnetic section
of the SpaCal with an azimuthal back-to-back topology.
Events with significant hadronic activity in the central region of
the detector were rejected if more than two tracks were found in the
Central Jet Chambers~\cite{Abt:1997hi}. The usage of this trigger was
possible, as in the analysed kinematic domain of low $y$
the final state hadrons are produced mostly at small polar angles.

The trigger efficiencies for both arrangements are determined using
independently triggered data.
The efficiency of the inelastic event selection by the second trigger
is evaluated as a function of the largest polar angle $\theta_{\rm LAr}$ of
all clusters in the LAr calorimeter which have an energy above
$0.5$\,GeV. It falls from $99\%$ at the minimum $\theta_{\rm LAr}$ possible
to $79\%$ at $\theta_{\rm LAr} = 30^\circ$.

For the event selection in the data analysis the following requirements
are imposed:
\begin{itemize}
\item {\bfseries Selection criteria for QED Compton events.}
The two most energetic clusters in the electromagnetic section of the SpaCal
are assumed to be produced by the scattered electron and the radiated photon.
The energy of each cluster must exceed $4$\,GeV, and the sum of both
energies must lie between $20$ and $30$\,GeV. The $e\gamma$ acoplanarity
defined via
$A = |180^\circ - \Delta\phi|$, where $\Delta\phi$ is the azimuthal angle
between the electron and photon directions, must be smaller than $45^\circ$.
Both particles must be found within the acceptance region of either the BST or
the CIP. The interaction vertex must be reconstructed, as described in
Sect.~\ref{s:detector}, within $30\,{\rm cm}$ in $z$ around the nominal
interaction point.
\item {\bfseries Additional selection criteria for the inelastic QEDC sample.}
At least one particle from the hadronic final state has to be detected
with energy above $0.5$\,GeV in the LAr calorimeter. The energy deposits below
$0.5$\,GeV are classified as noise.
The contribution from elastic QEDC events with noise in the LAr calorimeter
is further suppressed by requiring that the $e \gamma$ acoplanarity must
be larger than $2^\circ$. This requirement exploits the fact that the
acoplanarity is typically larger for inelastic events than for elastic events
due to the different $Q^2$ dependences of the two processes.
\item {\bfseries Background suppression requirements.}
The residual energy in the electromagnetic section of the SpaCal, given by
$E_{\rm res} = E_{\rm tot} - E_1 - E_2$, where $E_{\rm tot}$ is the total
energy and $E_1$, $E_2$ are the energies of the QEDC signal clusters, must be
below $1$\,GeV. This cut suppresses DIS, photoproduction, dielectron and 
vector meson backgrounds.
Furthermore, cuts on the shower shape estimators of the two SpaCal clusters
are imposed~\cite{Lendermann:2002} to separate electrons and photons from 
hadrons.
Finally, the phase space is restricted to $y_{\Sigma} < 0.0062$
(see eq.~(\ref{eq:kinrec})) to avoid kinematic regions with large contributions
from inclusive DIS background.
\end{itemize}
A brief summary of all selection criteria is given in Table\,\ref{tab:selection}.
After the selection 1938 events remain in the data sample.
\begin{table}[t]
\caption{Summary of QEDC selection criteria, as described in the text.}
\label{tab:selection}
\begin{center}
\begin{tabular}{@{}l@{\hspace{0.8cm}}l@{}}
\hline\noalign{\smallskip}
\multicolumn{1}{l}{\bfseries Purpose} & \multicolumn{1}{l}{\bfseries Criterion} \\
\hline\noalign{\smallskip}
  Inelastic QEDC signature  & \rule[-1.5mm]{0mm}{6mm} 
                              $E_1, E_2 > 4\,{\rm GeV}$                      \\
                            & \rule[-1.5mm]{0mm}{3mm} 
                              $20 < E_1 + E_2 < 30\,{\rm GeV}$               \\
                            & \rule[-1.5mm]{0mm}{3mm}
                              $2^{\circ} < A < 45^{\circ}$                   \\
                            & \rule[-1.5mm]{0mm}{3mm} 
                          At least one LAr cluster with $E > 0.5\,{\rm GeV}$ \\
                            & \rule[-1.5mm]{0mm}{3mm}
                              $|z_{\rm vtx}| < 30\,{\rm cm}$                 \\
  Background rejection      & \rule[-1.5mm]{0mm}{7mm} 
                            $E_{\rm res} < 1\,{\rm GeV}$                     \\
                            & \rule[-1.5mm]{0mm}{3mm}
                             Electromagnetic shower shape requirements       \\
                            & \rule[-1.5mm]{0mm}{3mm} 
                            $y_{\Sigma} < 0.0062$                            \\
\hline\noalign{\smallskip}
\end{tabular} 
\end{center}
\end{table}

\section{Event Kinematics and Control Distributions} \label{s:analysis}

The redundancy in the measured event properties permits the reconstruction 
of the variables $Q^2$, $x$ and $y$ either from the kinematics
of the scattered electron and radiated photon or from those
of the hadronic final state.
Due to the low $y$ values considered in the present analysis,
the variables $x$ and $y$ can, however,
not be determined solely from the measured electron and photon
four-momenta, since their resolution deteriorates as $1/y$.
Hence for the kinematic reconstruction the
$\Sigma$-method~\cite{Bassler:1995uq}
is employed, which uses information from the hadronic final state.
In addition, this method reduces the influence of higher order radiative
effects. With the proton beam energy $E_p$ as well as the sums
$\Sigma_h = \sum_i (E - p_z)_i$, where the summation
is performed over all hadronic final state objects, and
$\Sigma_{e\gamma} = (E - p_z)_e + (E - p_z)_\gamma$, this method yields
\begin{equation} \label{eq:kinrec}
y_\Sigma   = \frac{\Sigma_h}{\Sigma_h + \Sigma_{e\gamma}} ~,~~~~~~
Q^2_\Sigma = \frac{(\vec{p}_{t,e} + \vec{p}_{t,\gamma})^2}{1-y_\Sigma} ~,~~~~~~
x_\Sigma   = \frac{Q^2_\Sigma}{2 E_p \Sigma_h} ~.
\end{equation}
Here, the longitudinal ($p_z$) and transverse ($\vec{p}_t$) momenta
of the measured particles are calculated from their energies ($E$) and angles
neglecting particle masses.
As low $y$ values correspond to small polar angles of the final state hadrons,
one of the main challenges for the analysis is the correct
reconstruction of $\Sigma_h$ in light of the losses beyond the forward
acceptance of the detector.
This necessitates a good simulation of hadronisation processes at low
invariant masses $W$.

\begin{figure}[!tb]
\centerline{\includegraphics[width=\textwidth]{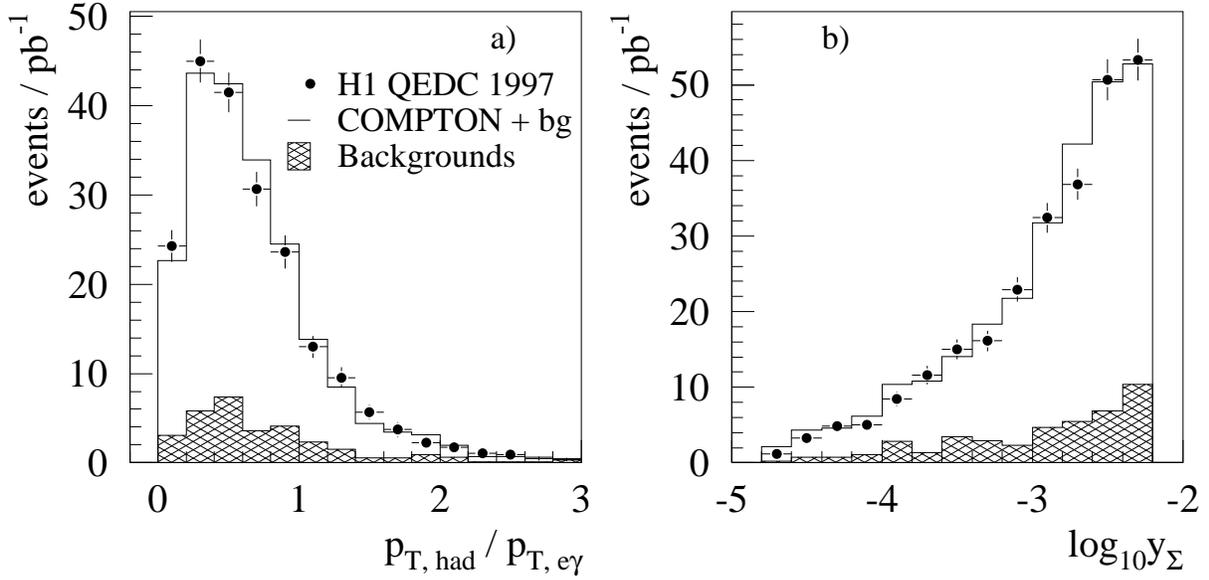}}
\caption{
a) ratio of the total measured transverse momentum of hadrons to the
total transverse momentum of the $e\gamma$ system;
b) uncorrected $y_\Sigma$ distribution after applying all selection cuts.
The data are depicted by the closed circles. 
The solid histogram represents the sum of COMPTON MC events and all
background contributions. The hatched histogram denotes the sum of
all background processes.}
\label{f:control1}
\end{figure}

The quality of the description of the hadronic final state, comprising 
both the generated hadronic distribution and
the subsequent simulation of the detector response, is illustrated in 
Fig.\,\ref{f:control1}a. The figure shows the ratio of the total transverse
momentum of measured hadrons, $p_{t,{\rm had}}$, to the total transverse
momentum of the $e\gamma$ system, $p_{t,e\gamma}$.
The simulation provides a very good description of the data, which is also
true for each phase space interval used in the $F_2$ measurement.
Due to losses in the very forward region, outside the acceptance of the LAr
calorimeter and the cut at $0.5$\,GeV on the LAr cluster energy,
the distribution peaks at values smaller than one. However, for the
calculation of the kinematic variables with the $\Sigma$-method, the total
$E - p_z$ of the hadrons is used, which is much less sensitive to losses in
the beam pipe than $p_{t,{\rm had}}$.

The control distribution of $y_\Sigma$, shown in Fig.\,\ref{f:control1}b,
demonstrates 
the good quality of both the hadronic simulation and the cross section
description given by the COMPTON program down to the lowest $y$ values,
even beyond the range used for the measurement. In order to compensate
the losses of hadronic energy and thus reduce migration effects,
a correction is applied to the measured $y$ which was determined from
the difference between the reconstructed and the generated $y$ 
in simulated events. The correction varies between $\sim$$20\%$ on average
in the lowest $y$ bins and $0$ at the highest $y$ values.

\begin{figure}[!tb]
\centerline{\includegraphics[width=\textwidth]{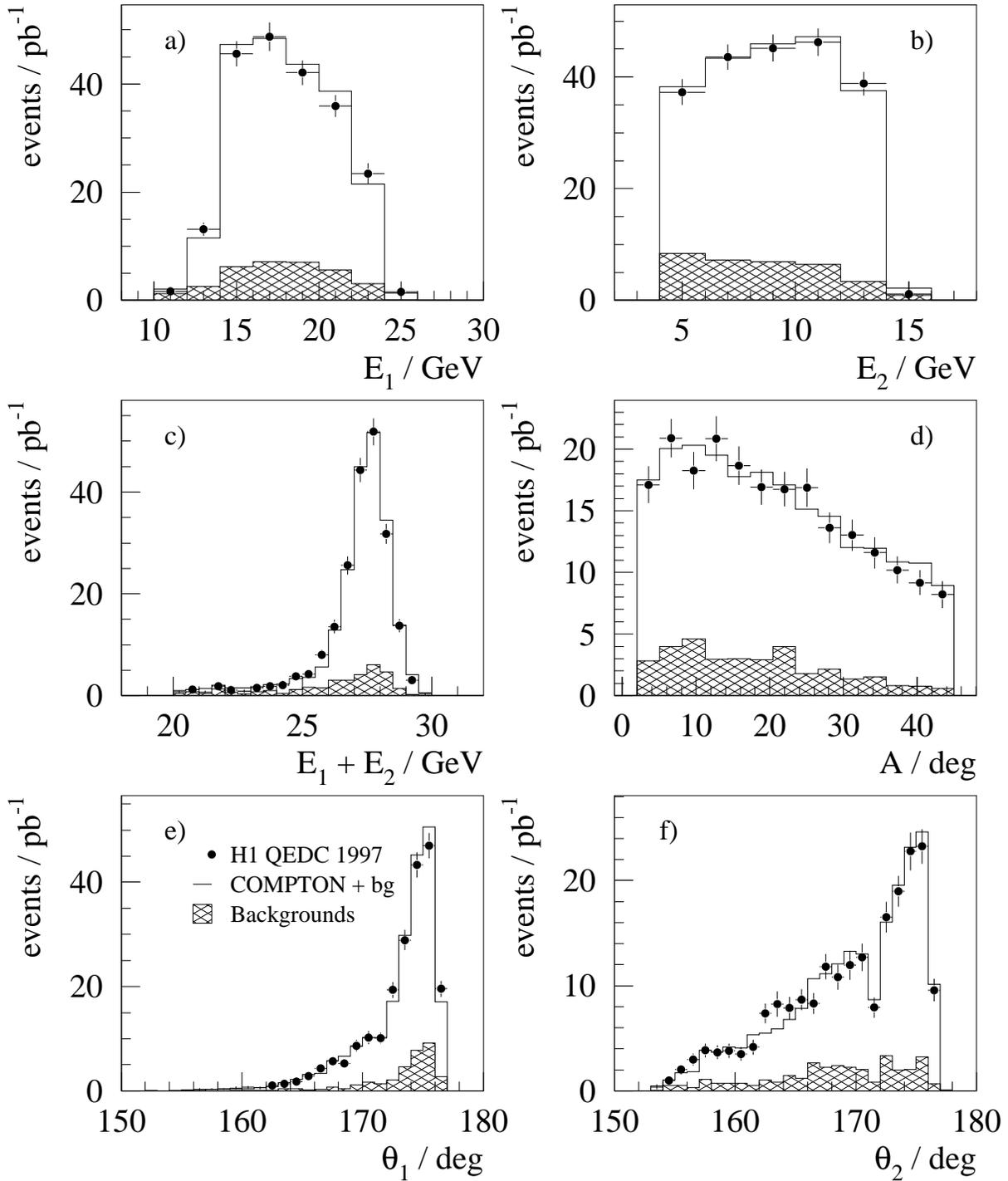}}
\caption{Control distributions for the measured electron and photon
in events used for the measurement:
a) energy of the particle with the higher energy;
b) energy of the particle with the lower energy;
c) sum of both energies; d) $e\gamma$ acoplanarity;
e) polar angle of the particle with the higher energy and
f) polar angle of the particle with the lower energy.
The data are depicted by the closed circles.
The solid histogram represents the sum of COMPTON MC events and all
background contributions. The hatched histogram denotes the sum of
all background processes.}
\label{f:control2}
\end{figure}

The control distributions in Fig.\,\ref{f:control2} illustrate the good
description of the electron-photon final state provided by the simulation.
The energies $E_1$, $E_2$ and the polar angles $\theta_1$, $\theta_2$
of the electron and the photon are shown (the subscripts 1 and 2 correspond
to the particle with higher and lower energy, respectively), as well as the
sum of both energies and the $e\gamma$ acoplanarity.
A dip in the polar angle distributions at $\sim$$172^{\circ}$ occurs due to
fiducial cuts imposed to ensure a precise description of the BST and CIP
acceptance.

\section{Results of the Measurement} \label{s:results}
In order to extract the structure function $F_2$ the data sample is divided
into subsamples corresponding to a grid in $y$ and $Q^{2}$. The bin sizes are
adapted to the resolution in the measured kinematic quantities such that the
stability and purity in all bins shown are greater than $30\%$. Here, 
the stability (purity) is defined as the ratio of the number of simulated
QEDC events originating from and reconstructed in a specific bin to the
number of generated (reconstructed) events in the same bin.

The measured value of the structure function at a given point in $Q^2$ and
$x=Q^2/ys$ is obtained from
\begin{equation}
F_2(x,Q^2) = \frac{N_{\rm data} - N_{\rm bg}}{N_{\rm MC}} \,
             F_2^{\rm MC}(x,Q^2) \;.
   \label{eq:f2extract}
\end{equation}
Here, $N_{\rm data}$ represents the number of selected events corrected
for the trigger efficiency (Sect.~\ref{s:selection}),
$N_{\rm bg}$ is the sum of all background contributions, estimated as
described in Sect.~\ref{s:background}, $N_{\rm MC}$ is the number of
reconstructed events in the COMPTON simulation rescaled to the luminosity of
the data, and $F_2^{\rm MC}$ is the structure function value given by the
parameterisation used in the COMPTON simulation. This method takes into
account bin-to-bin migration effects, as well as bin-centre and higher order
radiative corrections.

Systematic uncertainties are determined by studying the stability of the
results under variations of the measured energies and angles, 
selection efficiencies and background normalisations
as well as modifications of the theoretical input.
The following systematic errors are estimated:
\begin{itemize}
\item The uncertainty on the hadron measurement, which
  takes into account the errors of the hadronic final state simulation,
  detector acceptance and LAr calorimeter energy scale, is estimated
  from the comparison between the data and the simulation in the $p_t$ balance
  between the reconstructed hadron and electron-photon final states. The 
  contribution to the systematic error varies between $1$ and $8\%$;
\item
  A $10\%$ uncertainty on the energy attributed to noise
  in the LAr calorimeter is estimated by studying the quality of the simulation
  of the energy depositions rejected by the noise cut.
  The contribution to the systematic error varies from $1.5\%$
  at the highest $y$ values to $12.5\%$ at the lowest $y$;
\item The uncertainty on the SpaCal energy scale, affecting
  simultaneously the energies of the electron and the photon, amounts to
  $0.5\%$ at $E > 17$\,GeV and increases linearly towards lower energies 
  up to $2.7\%$ at $E = 4$\,GeV. This source yields a $1 - 2\%$ error
  on the $F_2$ measurement;
\item The errors on the $F_2$ measurement due to the uncertainties in
  the electron and photon polar angle reconstruction ($0.3 - 2.5$\,mrad)
  are typically below $1\%$;
\item The efficiency of the vertex
reconstruction is studied using inclusive DIS and elastic QEDC events.
The corresponding error contribution to the $F_2$ measurement is estimated
to be $2\%$;
\item Amongst the errors from the background estimation, described 
in Sect.~\ref{s:background}, the main contributions arise from inelastic DVCS
($5.5\%$) and inclusive DIS (up to $6\%$). The errors from other contributions
are $\lesssim 1\%$;
\item The statistical errors from the signal and background Monte Carlo event
samples contribute up to $5.5\%$ uncertainties on the measurement;
\item The luminosity measurement contributes a global $1.5\%$ error;
\item The uncertainties on the trigger efficiencies are all below $0.5\%$.
In total they contribute $\lesssim 1\%$ error to the measurement;
\item The uncertainty due to radiative corrections, as simulated by the
COMPTON generator, amounts to $2\%$.
\end{itemize}
The statistical errors lie in the range $6 - 10\%$, while the systematic
uncertainties are typically $9 - 12\%$, rising to $18\%$ at the lowest
$y$ values. The total errors are obtained by adding the statistical and
systematic errors in quadrature.

\begin{figure}[!p]
\centerline{\includegraphics[width=\textwidth]{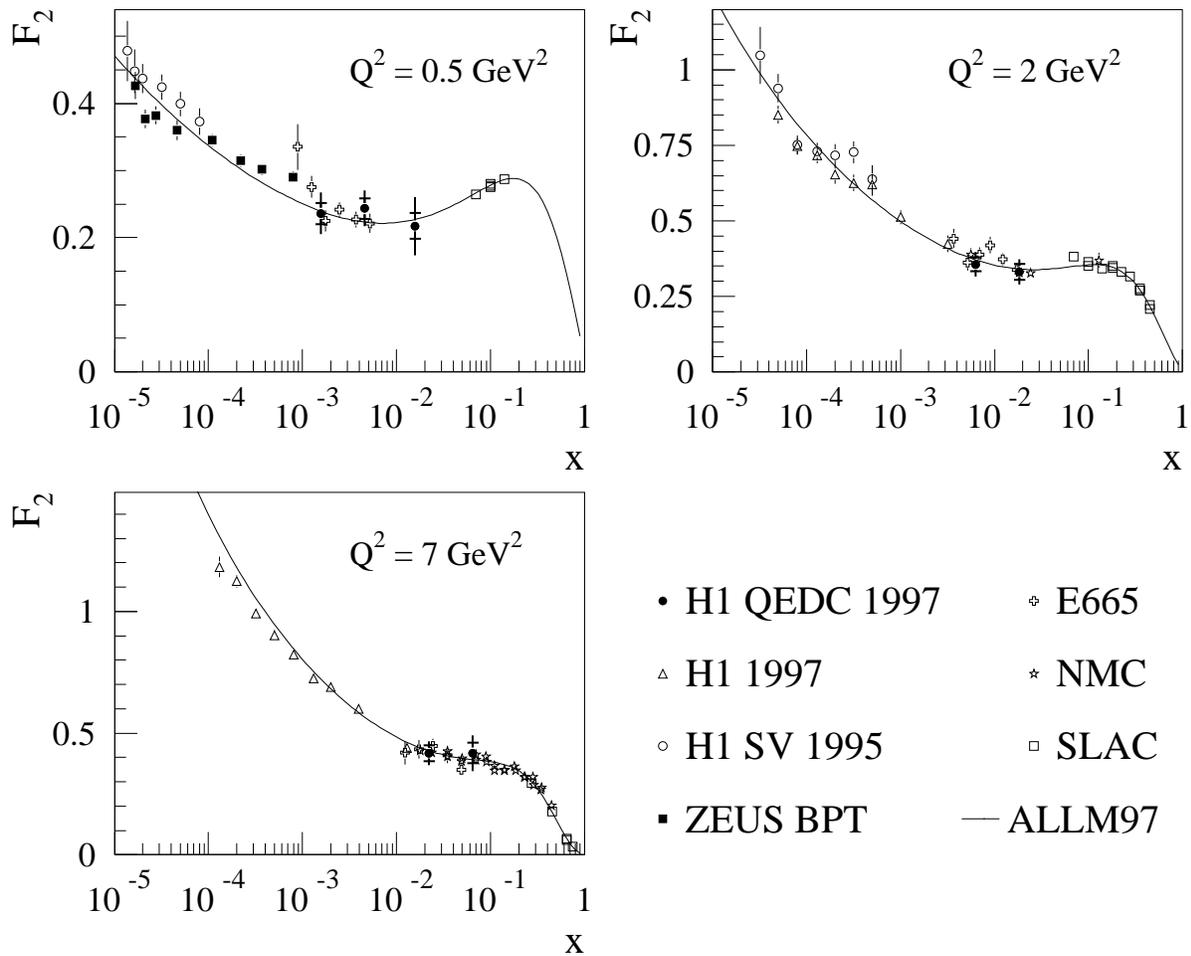}}
\caption{$F_2$ measurements from QED Compton scattering by H1
(closed circles), compared with other measurements at HERA
(closed squares~\cite{Breitweg:2000yn}, open triangles~\cite{Adloff:2000qk}
and open circles~\cite{Adloff:1997mf}) and fixed target experiments
(open squares~\cite{Whitlow:1992uw}, open stars~\cite{Arneodo:1997qe}
and open crosses~\cite{Adams:1996gu}). The inner error bars for the
QEDC data represent the statistical errors and the total error bars
the statistical and systematic errors added in quadrature.
The solid line depicts the ALLM97 parameterisation~\cite{Abramowicz:1997ms}.
The data of the other measurements are shifted to the $Q^2$ values of the
present measurement using the ALLM97 parameterisation.}
\label{f:f2qedc}
\end{figure}

The $F_2$ values measured in QED Compton scattering, as summarised in Table\,%
\ref{t:results}, are shown in Fig.\,\ref{f:f2qedc} as a function of $x$
at fixed $Q^2$ and are compared with other HERA~%
\cite{Adloff:1997mf,Adloff:2000qk,Breitweg:2000yn}
and fixed target~\cite{Whitlow:1992uw,Arneodo:1997qe,Adams:1996gu} data.
The present analysis extends the kinematic range of HERA measurements at low
$Q^2$ towards higher $x$ values, thus complementing standard inclusive
and shifted vertex measurements.
The region covered overlaps with the domain of fixed target experiments.
The QEDC $F_2$ data are consistent with their results. The data are also well
described by the ALLM97 parameterisation~\cite{Abramowicz:1997ms}.

\section{Summary}  \label{s:conclusion}
The first measurement of the proton structure function $F_2$ using QED Compton
scattering at HERA is presented. The available range of phase space extends
down to $Q^2 = 0.5\,{\rm GeV}^2$, in the transition
region from DIS to photoproduction. Due to an improved treatment of the
hadronic final state at small masses $W$, the measurement extends the low $Q^2$
kinematic domain of HERA up to $x \sim 0.06$, complementing the analyses of
standard inclusive DIS data and data taken with the interaction vertex shifted.
The results are in good agreement with data from fixed target 
lepton-nucleon scattering experiments.

\section*{Acknowledgements}
\refstepcounter{pdfadd} \pdfbookmark[0]{Acknowledgements}{s:acknowledge}
We are grateful to the HERA machine group whose outstanding
efforts have made this experiment possible. 
We thank the engineers and technicians for their work in constructing
and maintaining the H1 detector, our funding agencies for 
financial support, the
DESY technical staff for continual assistance
and the DESY directorate for support and for the
hospitality which they extend to the non DESY 
members of the collaboration.

\clearpage

\begin{table}[!tbp]
\tablecaption{Measurements of the proton structure function $F_2(x,Q^2)$
in QED Compton scattering, together with fractional 
statistical~($\delta_{\rm stat}$), systematic~($\delta_{\rm syst}$) and
total errors~($\delta_{\rm tot}$). The $F_2$ values
at the points of $Q^2$, $x$, $y$ are extracted from the event rates
measured in the kinematic bins $\Delta Q^2$, $\Delta y$.}
\centerline{%
\begin{tabular}{|@{ }c@{\hspace*{0.1cm}}c@{\hspace*{0.4cm}}c@{\hspace*{0.2cm}}|@{\hspace*{0.2cm}}r@{ -- }r@{\hspace*{0.4cm}}r@{ -- }r@{\hspace*{0.2cm}}|@{\hspace*{0.2cm}}c@{\hspace*{0.2cm}}r@{\hspace*{0.4cm}}r@{\hspace*{0.4cm}}r@{ }|}
\hline
\rule[-1.5ex]{0ex}{4ex}$Q^2$,~GeV$^2$ & $x$ & $y$ & 
\multicolumn{2}{@{}l@{}}{$\Delta Q^2$,~GeV$^2$} & 
\multicolumn{2}{@{}c|@{\hspace*{0.2cm}}}{$\Delta y$} & $F_2$ & 
\multicolumn{1}{@{}c@{\hspace*{0.3cm}}}{$\delta_{\rm stat}$,\%} &
\multicolumn{1}{@{}c@{\hspace*{0.3cm}}}{$\delta_{\rm syst}$,\%} &
\multicolumn{1}{@{}c@{ }|}{$\delta_{\rm tot}$,\%} \\
\hline
\rule{0ex}{2.5ex}%
0.5 & 0.01578 & 0.00035 & 0.1 &  1.0 & 0.00020 & 0.00062 & 0.217 &  8.8 & 17.9 & 19.9 \\
0.5 & 0.00460 & 0.00120 & 0.1 &  1.0 & 0.00062 & 0.00200 & 0.243 &  6.3 &  9.0 & 11.0 \\
0.5 & 0.00158 & 0.00350 & 0.1 &  1.0 & 0.00200 & 0.00620 & 0.236 &  6.7 & 11.4 & 13.2 \\
2.0 & 0.01841 & 0.00120 & 1.0 &  3.0 & 0.00062 & 0.00200 & 0.331 &  7.7 & 10.0 & 12.6 \\
2.0 & 0.00631 & 0.00350 & 1.0 &  3.0 & 0.00200 & 0.00620 & 0.356 &  6.3 &  9.3 & 11.2 \\
7.0 & 0.06444 & 0.00120 & 3.0 & 20.0 & 0.00062 & 0.00200 & 0.418 &  9.9 & 14.5 & 17.6 \\
7.0 & 0.02209 & 0.00350 & 3.0 & 20.0 & 0.00200 & 0.00620 & 0.418 &  7.5 &  8.9 & 11.7 \\
\hline
\end{tabular}}
\label{t:results}
\end{table}

\end{document}